\documentclass[10pt,twocolumn,letterpaper]{article}

\usepackage{graphicx}
\usepackage{amsmath}
\usepackage{amssymb}
\usepackage{booktabs}

\usepackage[utf8]{inputenc} 
\usepackage[T1]{fontenc}    
\usepackage{hyperref}       
\usepackage{url}            
\usepackage{booktabs}       
\usepackage{amsfonts}       
\usepackage{nicefrac}       
\usepackage{microtype}      

\usepackage{multirow}
\usepackage{array}
\usepackage{graphbox}
\usepackage{color}
\usepackage{graphicx}
\usepackage{comment}
\usepackage{amsmath,amssymb} 
\usepackage[normalem]{ulem}
\usepackage{lipsum}  

\usepackage[ruled]{algorithm2e} 

\usepackage[capitalize]{cleveref}
\crefname{section}{Sec.}{Secs.}
\Crefname{section}{Section}{Sections}
\Crefname{table}{Table}{Tables}
\crefname{table}{Tab.}{Tabs.}

\newcommand*{\ShowNotes}{}

\definecolor{darkred}{rgb}{0.7,0.1,0.1}
\definecolor{darkgreen}{rgb}{0.1,0.7,0.1}
\definecolor{cyan}{rgb}{0.7,0.0,0.7}
\definecolor{dblue}{rgb}{0.2,0.2,0.8}
\definecolor{maroon}{rgb}{0.76,.13,.28}
\definecolor{burntorange}{rgb}{0.81,.33,0}

\ifdefined\ShowNotes
  \newcommand{\colornote}[3]{{\color{#1}\bf{#2: #3}\normalfont}}
\else
  \newcommand{\colornote}[3]{}
\fi

\begin{document}

\title{Random Walks in Self-supervised Learning for Triangular Meshes}

\author{Gal Yefet\\
Technion – Israel Institute of Technology\\
Israel\\
{\tt\small galyefet@campus.technion.ac.il}
\and
Ayellet Tal\\
Technion – Israel Institute of Technology\\
Israel\\
{\tt\small ayellet@ee.technion.ac.il}
}
\maketitle

\begin{abstract}
This study addresses the challenge of self-supervised learning for 3D mesh analysis.
It presents an new approach that uses random walks as a  form of data augmentation to generate diverse representations of mesh surfaces.
Furthermore, it employs a combination of contrastive and clustering losses. 
The contrastive learning framework maximizes similarity between augmented instances of the same mesh while minimizing similarity between different meshes. 
We integrate this with a clustering loss, enhancing class distinction across training epochs and mitigating training variance. 
Our model's effectiveness is evaluated using mean Average Precision (mAP) scores and a supervised SVM linear classifier on extracted features, demonstrating its potential for various downstream tasks such as object classification and shape retrieval. 

\end{abstract}

\section{Introduction}

\begin{figure*}
\begin{center}
    \begin{tabular}{c|c c c}
         Query &  \#1 & \#2 & \#3 \\
         \includegraphics[width=0.15\textwidth]{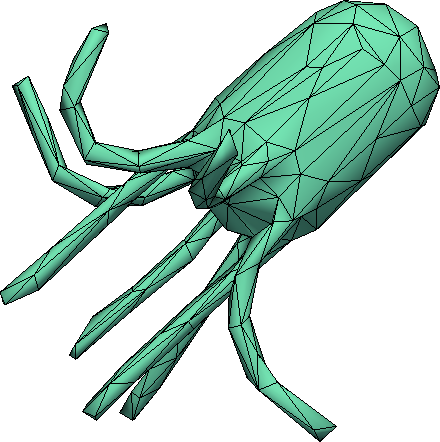}&  \includegraphics[width=0.15\textwidth]{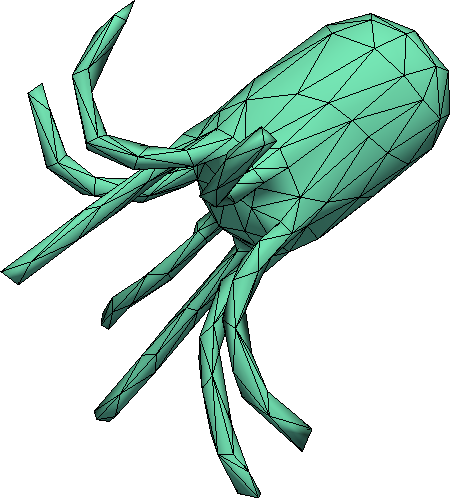}&  \includegraphics[width=0.15\textwidth]{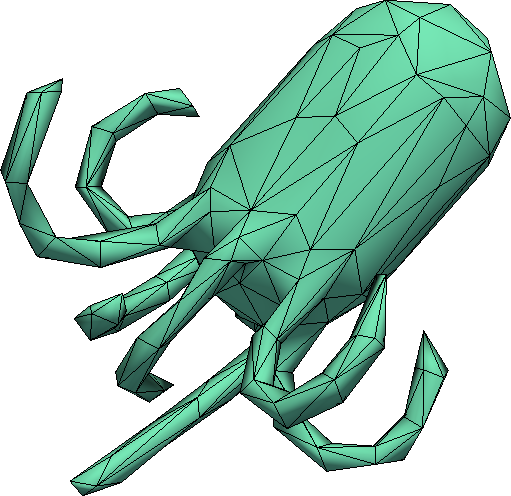}&  \includegraphics[width=0.15\textwidth]{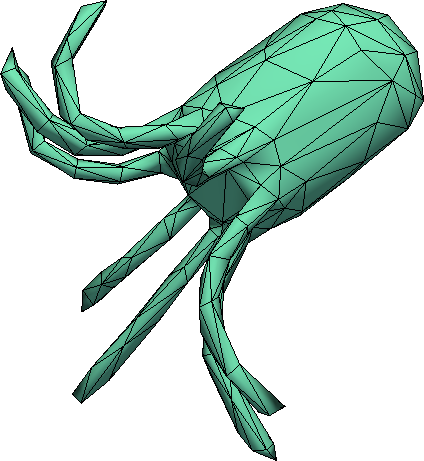}\\
         \hline
         \includegraphics[width=0.15\textwidth]{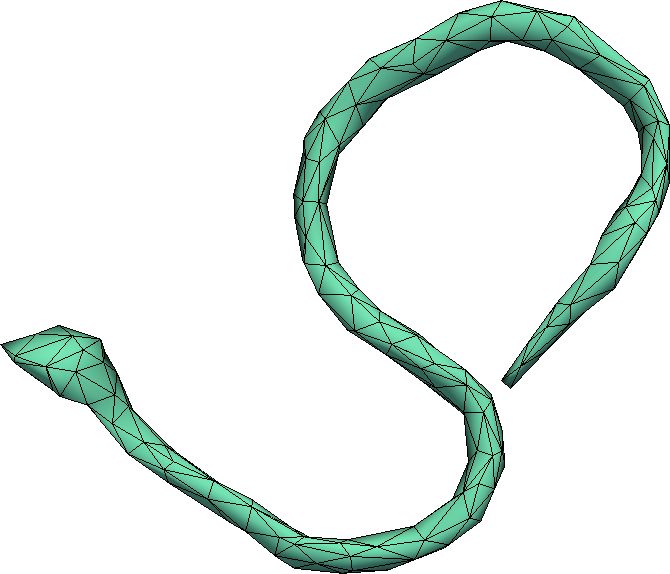}&  \includegraphics[width=0.15\textwidth]{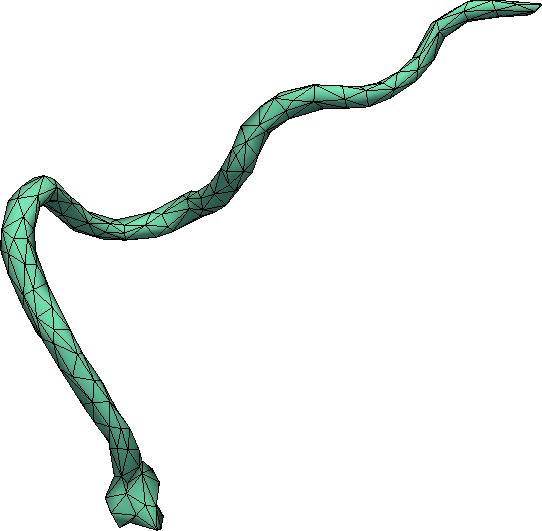}&  \includegraphics[width=0.15\textwidth]{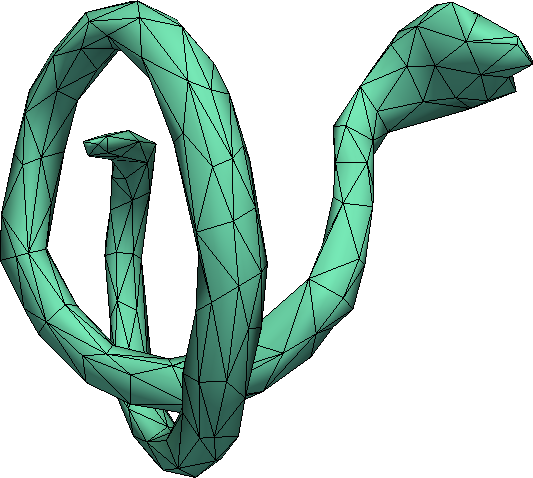}&  \includegraphics[width=0.15\textwidth]{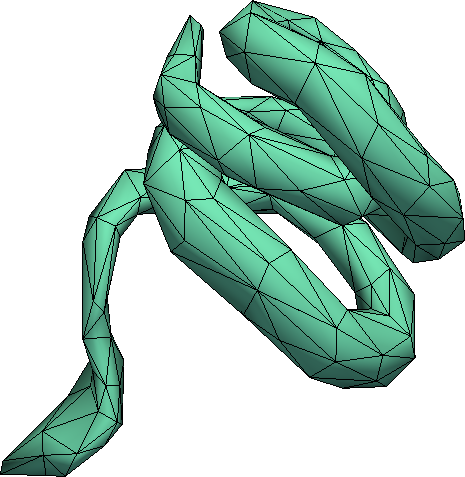}\\
         \hline
         \includegraphics[width=0.15\textwidth]{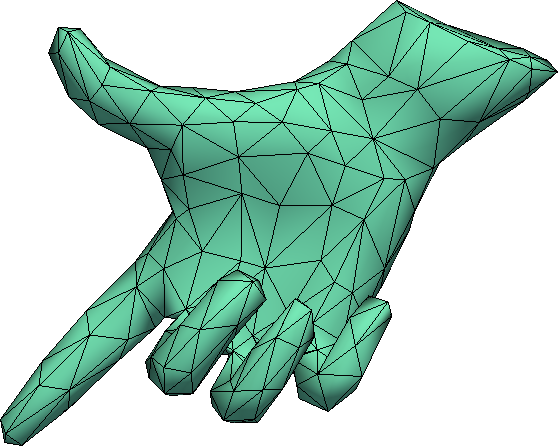}&  \includegraphics[width=0.15\textwidth]{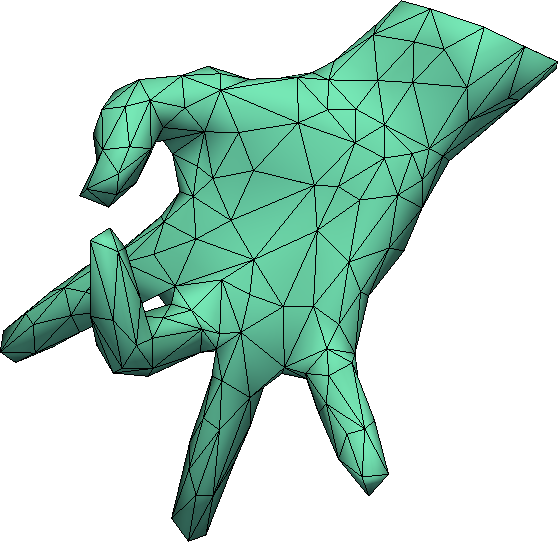}&  \includegraphics[width=0.15\textwidth]{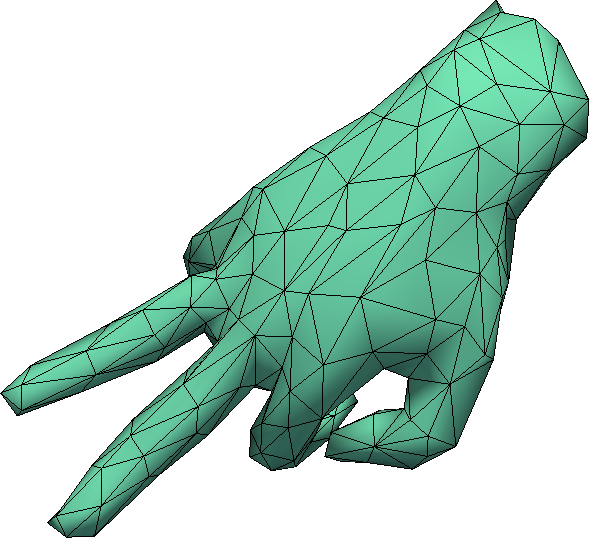}&  \includegraphics[width=0.15\textwidth]{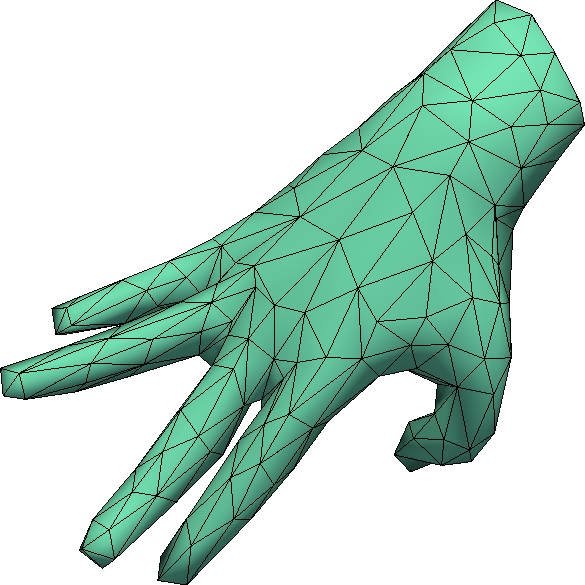}\\
         \hline  
    \end{tabular}
  \caption{{\bf Model Retrieval Results for the SHREC11 Dataset~\cite{veltkamp2011shrec}}
This figure demonstrates the retrieval performance of our model on three query examples from the SHREC11 dataset: an octopus, a snake, and a hand. The results showcase the model's ability to successfully retrieve similar items in the same class, even when the objects are presented in various poses or orientations~\cite{veltkamp2011shrec}. }
\label{fig:retrieval_shrec11_intro}
\end{center}
\end{figure*}

Shape analysis of three-dimensional (3D) objects plays a crucial role in contemporary research within the fields of computer vision and computer graphics. Its significance stems from its application across various domains, such as self-driving cars, virtual and augmented reality, robotics, and medicine, among others. Numerous representations exist for 3D objects, with triangular meshes, point clouds, and volumetric data being the most prominent. 

This study specifically concentrates on triangular meshes, as they are widely used in computer graphics due to their efficiency and high quality. However, 3D meshes possess unordered and irregular characteristics, which pose challenges for deep learning algorithms. Consequently, efforts have been made to rearrange the data and redefine the convolution and pooling operations in order to enable the utilization of convolutional neural networks (CNNs) ~\cite{atzmon2018point, hua2018pointwise, li2018pointcnn, thomas2019kpconv, xu2018spidercnn}.

Our research takes inspiration from a recent framework called Meshwalker~\cite{lahav2020meshwalker}, which takes a different approach.
Instead of utilizing {\em convolutional neural networks (CNNs)}, Meshwalker suggests capturing the geometry and topology of a mesh by employing random walks along its surface. Each walk is processed by a {\em Recurrent Neural Network (RNN)}~\cite{cho2014learning,journals/pami/GravesLFBBS09,hochreiter1997long},  which gathers surface information throughout the walk. Multiple random walks, generated independently, represent a given mesh. Meshwalker achieves impressive results in supervised learning, showcasing the effectiveness of randomness.

This research, however, focuses on the domain of self-supervised learning. This is particularly relevant due to the difficulty of  manual labeling of 3D data. Self-supervised learning aims to uncover patterns and extract meaningful features from data without relying on explicit labels. These extracted features can be leveraged for various purposes such as clustering, dimensionality reduction, or as a pre-training step for subsequent tasks. In the realm of two-dimensional (2D) data, self-supervised learning has been extensively explored~\cite{chen2020simple, henaff2020data, hjelm2018learning, oord2018representation, tian2020contrastive,wu2018unsupervised, caron2020unsupervised}. In the context of three-dimensional (3D) data, several works have focused on point clouds~\cite{sharma2016vconv, wu2016learning, achlioptas2018learning, yang2018foldingnet, li2020unsupervised}. The underlying idea involves creating augmented versions of 3D objects by applying transformations such as rotation, scaling, noise addition, etc. These augmented versions, known to be related to the same object, are then used to train the network. Some approaches employ contrastive loss~\cite{li2020unsupervised, xie2020pointcontrast, jiang2021unsupervised}, while others utilize an auto-encoder framework~\cite{sharma2016vconv, wu2016learning, achlioptas2018learning, yang2018foldingnet}.

Our focus, however, is on triangular meshes.
Meshes are able to capture detailed local surface properties, due to the explicit  neighborhoods and connectivity. 
While the absence of point connectivity may not pose a problem in certain conditions, there are cases in which it becomes exceedingly difficult to deduce any meaningful information from the point cloud object, as highlighted in the case of the Engraved Cubes dataset of~\cite{Hanocka_2019}.

Our key idea is to address the challenges of self-supervised learning in the context of meshes by utilizing random walks as augmentations.
This may serve as a powerful mechanism to effectively represent the same model with different features. 
By performing random walks on both the same model and other models, we employ contrastive loss to encourage the network to learn meaningful representations for each class. This approach leverages the descriptive nature of random walks as augmentations, enabling the network to capture and characterize the intricacies of the same model more comprehensively.

For the contrastive loss, we employ the Normalized Temperature-scaled Cross Entropy Loss (NT-Xnet), which was initially introduced in SimCLR~\cite{chen2020simple}. NT-Xnet loss aims to maximize the similarity between augmented instances produced from the same mesh while simultaneously minimizing the similarity between instances derived from different meshes. This prompts the network to discern fine-grained differences between classes.
The random walks provide a mechanism to capture the local geometric context, while the contrastive loss leverages the relationships between the random walks, to guide the learning process. This synergy allows our solution to uncover meaningful patterns and features that can be harnessed for various downstream tasks, such as object classification, segmentation, or shape retrieval.

An additional crucial component in our architecture involves the integration of a clustering loss, which serves to enhance the notion of classes across different epochs. This approach aims to mitigate training variance by seeking proximity to the consensus established in previous epochs, rather than solely focusing on fitting to the agreement of the current epoch. The inclusion of clustering loss, alongside contrastive loss, has been explored and showed great potential in the 2D field~\cite{yan2019clusterfit, caron2021unsupervised}.

Our model was evaluated using two self-supervised shape analysis applications: retrieval and classification. In the latter, we use a linear SVM classifier applied to the extracted features.
%
We trained and tested our model using two commonly-used datasets: SHREC11~\cite{veltkamp2011shrec} and ModelNet40~\cite{wu20153d}. 
These datasets offer diverse examples to evaluate our model's performance in learning discriminative representations for 3D shape tasks. 

{Figure~\ref{fig:retrieval_shrec11_intro} illustrates the effectiveness of the model's retrieval capabilities using three sample queries from the SHREC11 dataset: an octopus, a snake, and a hand. The displayed results highlight the model's proficiency in identifying and retrieving objects of the same class as the query, demonstrating its robustness in handling diverse poses and orientations within each object category.

This work makes three contributions:
\begin{enumerate}
\item The introduction of a novel self-supervised learning framework for 3D meshes.
\item The utilization of random walks as an augmentation for feature extraction in a self-supervised settings.
\item Achieving classification accuracy on the SHREC11 dataset that is only 2\% lower than a supervised model demonstrates the effectiveness of our self-supervised approach in 3D shape analysis.
\end{enumerate}
These contributions highlight the potential of self-supervised learning techniques in advancing the understanding and analysis of 3D meshes.

\section{Related work}
\label{sec:related}

\subsection{Mesh analysis}
A triangular mesh, consisting of vertices (V), edges (E), and faces (F), is the most common and effective 3D shape representation in computer graphics. 
Despite its popularity, the irregular and unstructured nature of this representation presents challenges for applying Convolutional Neural Networks (CNNs), as each vertex can have a varying number of neighbors at different distances. 

Traditional deep learning methods address this irregularity by converting meshes into volumetric grids \cite{Maturana2015VoxNet, Muzahid2020Octree, Qi2016Volumetric} or creating multiple 2D projections \cite{Kanezaki2018RotationNet} to make them compatible with CNNs. 
Another approach involves extracting 3D point samples from the surface of an object and using point cloud techniques to analyze these samples, employing either multi-layer perceptrons or convolutional layers to learn representations based on neighboring points \cite{Qi2017PointNet, Qi2017PointNetPlusPlus, Sun2019SRINet, Tao2021Multihead, Wang2019DeepGL, Zhang2019RotationInvariant}. 

Recently, there has been an increasing interest in directly addressing 3D meshes. 
Earlier research focused on adapting 2D CNNs to 3D by treating meshes as manifolds and applying patch-based techniques on their surfaces \cite{Lim2018IntrinsicCorrespondence, Masci2015GeodesicCNN, Monti2017Geometric, Hanocka_2019, li2023tpnet}, or by using spatial diffusion as the main network operation \cite{sharp2022diffusionnet}. 
However, these patch-based methods often involve high computational complexity, hand-crafted techniques, and require precomputed local coordinate systems.
Some other recent methods focused on using graphs to enhance the rotation-invariant capabilities \cite{shakibajahromi2024rimeshgnn} and use transformers to grasp global attention \cite{vecchio2023metgraphtransformersemantic}.
Another approach to managing the irregular structure of 3D meshes involves using multiple random walks to explore the mesh's surface, which allows for the examination of both its local and global geometry \cite{lahav2020meshwalker, Izhak2022AttWalk}. 
These random walks are then fed into a Recurrent Neural Network (RNN), capable of retaining each walk's history.
Our work is influenced by this approach.

\subsection{Self-supervised learning}
Self-supervised learning involves creating pretext tasks directly from the data, which are then used to pre-train the model. 
In computer vision, these pretext tasks include predicting the sequence of events over time \cite{wei2018learning}, identifying missing pixels \cite{pathak2016context}, locating patches \cite{Doersch2015Unsupervised}, determining image orientations \cite{Gidaris2018Unsupervised}, recognizing human-made artifacts \cite{Jenni2018SelfSupervised}, clustering images \cite{Caron2018DeepClustering}, identifying camera positions \cite{Agrawal2015Learning}, solving jigsaw puzzles \cite{Noroozi2016Unsupervised}, predicting video colors \cite{Vondrick2018Tracking}, and tracking image patches \cite{Wang2015Unsupervised}. 
These approaches have shown promising results in transferring visual features from pretext tasks to other tasks, emphasizing the importance of defining pretext tasks that are closely related to the downstream task \cite{Jenni2018SelfSupervised}.

In the area of 3D data, self-supervised learning has primarily focused on point clouds. Techniques include multitask learning \cite{Hassani2019Unsupervised}, reconstruction \cite{achlioptas2018learning}, contrastive learning \cite{Zhang2019Unsupervised}, restoring point clouds \cite{Shi2020Unsupervised}, point cloud autoregression \cite{Sun2020PointGrow}, multi-modal
\cite{crosspoint}, orientation prediction 
\cite{Han2019MultiAngle}, and approximating convex decomposition \cite{Gadelha2020LabelEfficient} to pre-train models and achieve state-of-the-art results in point cloud classification and segmentation tasks. 

Recently, masked autoencoders, which reconstruct data from masked inputs, have gained popularity in self-supervised learning, especially for images \cite{He2022Masked} and graphs \cite{Zhang2022GraphMAE}.
Other recent research has employed a multi-modal approach, combining 3D models and corresponding 2D images for training \cite{crosspoint, hao2024bootstrapping}.

\section{Model}
\label{sec:model}

\begin{figure*}[t]
\centering
\begin{tabular}{c}
\includegraphics[width=0.8\textwidth]{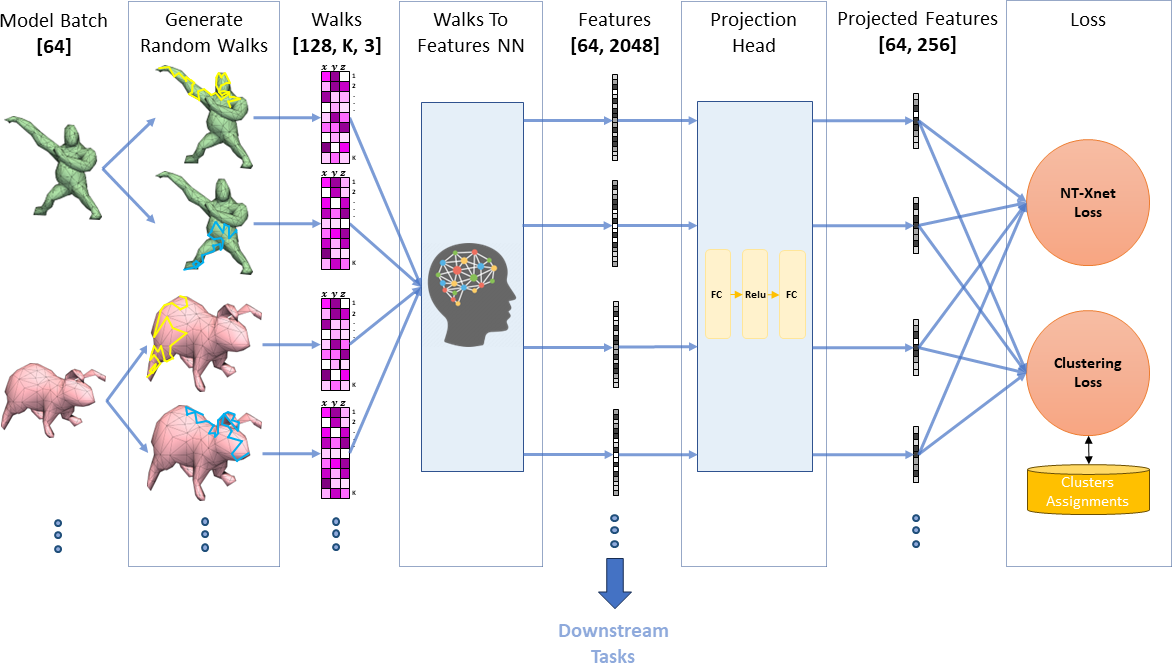}\end{tabular}
\caption{{\bf Architecture.} 
The model consists of four components: 
The first component is responsible for generating the walks from the 3D mesh models. The second component aggregates the information along the walk and yields the features. The third component is the projection head, which applies non-linear fully connected layers to the features in order to map input data into a lower-dimensional space, aiding in learning meaningful representations. 
The last component comprises the loss functions (a combination of the NT-Xnet loss and a Clustering loss). 
The outcome features that will be used for a downstream task are the features before the projection head.
}
\label{fig:arch}
\end{figure*}

We are provided with a collection of mesh objects in a dataset, and our model's task is to identify and extract features that are distinctive to the unknown classes from which these objects originate.
Our model realizes the following key idea: The embeddings of random walks from the same class should reside next to each other. Conversely, the embeddings of random walks from different classes should be farther apart. The problem, however, is that we do not know which objects belong to the same class. To address this, we pull together walks from the same model and push apart walks from different models. However, this does not guarantee that different objects from the same class will be clustered together. We propose to encourage this by introducing a new loss function, the {
\em clustering loss,} which encourages embeddings to form clusters.


Figure~\ref{fig:arch} illustrates our proposed architecture aimed at self-supervised learning. 
At first, inline with the MeshWalker~\cite{lahav2020meshwalker} approach, we create two random walks for every mesh object.
A walk is a sequence of $k$ vertices (steps), each of which is represented by its $X,Y,Z$ coordinates.
Once we have collected these walks from each model into a batch, they're fed into our {\em Walks-to-Features} neural network. 
This network's role is to transform these sequential random walks into a feature space; these features will be used for evaluation and for downstream tasks. 
Next, during the training phase, theses features are processed through an MLP non-linear projection head. 
This step condenses the features into a smaller, more manageable dimension, which has been proven to be essential for training~\cite{chen2020simple}.
Our losses are then computed on these features in the batch.
The two loss functions we apply are the {\em NT-Xnet contrastive loss} and a {\em clustering loss,} both of which enable the model's learning of the most significant attributes for each class.


\begin{figure}[t]
\centering
\includegraphics[width=0.45\textwidth]{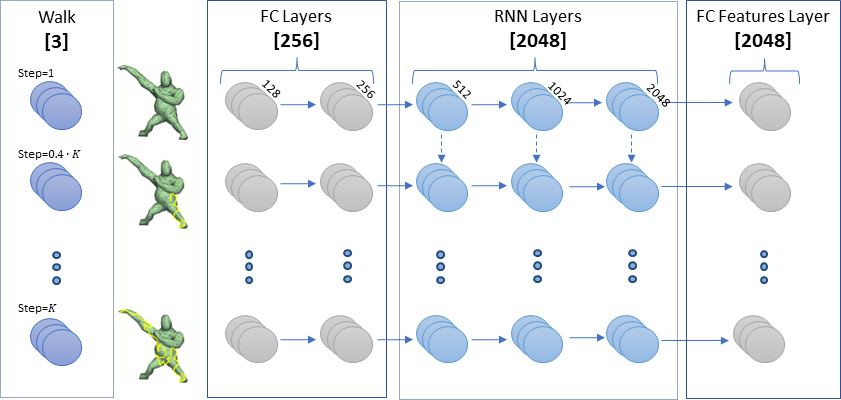}
\caption{ {\bf Walk To Features architecture~\cite{lahav2020meshwalker}.}
This NN  is initiated with a batch of random walks (the yellow walk on the gorilla). 
It consists of $3$ components:
(1) The FC layers change the feature space; 
(2) The RNN layers aggregate the information along the walk; 
(3) The FC layer gathers the features.
} 
\label{fig:meshwalker}
\end{figure}

Our model consists of two parts. 
The first part, aiming to generate walks and embed them in space, follows~\cite{lahav2020meshwalker} (Section~\ref{subsec:walk}). 
The second part learns to project features, adjusting the proximity of walks based on our key idea (Section~\ref{subsec:proj}). 
In the following, we will elaborate on these components. 
We are provided with a dataset containing a diverse set of mesh objects, with each object belonging to an unknown class.

\subsection{Creating walk embedding}
\label{subsec:walk}


\noindent
{\bf Data preparation.}
Our goal is to simplify the models for more effective walks and generate augmentations that are guaranteed to be from the same class. 
We apply classical approaches to downscale or upscale each model to augmentations with $1K$, $2K$, and $4K$ faces. 
This process resembles resizing an image, contributing to utilizing a uniform walk length across all models.
Importantly, as we generate these variations, we retain the source model ID for each version. 
This assists subsequent stages in establishing connections between two distinct models. 

{\bf Random walk generation.}
Given a batch of meshes ($64$ in our implementation), our goal is to create a batch of walks from which we will be able to extract qualitative features, to differentiate between various classes.
Toward this end, we propose to create two random walks for each model, both sharing the same {\em model ID}, so that we know for certain that their features should be close, even though the walks can be very different.

Similarly to~\cite{lahav2020meshwalker}, we start each walk at a random vertex of the mesh and sample its $\{X,Y,Z\}$ values. 
Subsequently, we move randomly to a neighboring vertex, occasionally jumping to another random vertex based on a small probability. 
We repeat this process for $K$ steps.

{\bf Walk-to-Features neural network}
We are given a batch of walks, and aim at extracting qualitative features for each walk.
We follow~\cite{lahav2020meshwalker} and employ a {\em Recurrent Neural Network (RNN)} to "remember" and accumulate knowledge throughout the walk.
As illustrated in Figure~\ref{fig:meshwalker},  the first sub-network consists of Fully Connected (FC) layers to convert the 3D input feature space into $256$-dimensional feature space. 
The second sub-network is the core part, consisting of RNN layers. 
These three layers are fed a sequence of walk steps
and output an accumulated $2048$ features vector. 
Finally, the last sub-network involves a single FC layer to utilize the RNN's output features in creating the final features vector. 

\subsection{Features  and Losses}
\label{subsec:proj}

At this stage, we have extracted the features of the walks, and we proceed to encourage features of models from the same class to converge, while moving apart features from other classes. 
To accomplish this, we employed a contrastive loss and a new clustering loss, Which complement each other as we will see below.
They are applied after a non-linear projection head, which enhances the contrastive performance.

{\bf Projection Head.}
The goal of employing a projection head before implementing a contrastive loss is to create a denser feature space, which will be used for the contrastive loss. 
Previous findings, as indicated by~\cite{chen2020simple}, have demonstrated increased effectiveness of the contrastive loss on denser features, even when the target features precede the projection head. 
Applying the projection head removes information that might be useful for downstream tasks but is less crucial for the contrastive loss objective, such as the orientation of the model.
In particular, we are given a batch of 64 feature vectors consisting of 2048 elements each and our target is to compress them into 256 elements. 
To achieve this, we utilized a simple non-linear projection head by applying two fully connected layers with a ReLU activation function. 

{\bf Loss Functions.}
Our aim is to apply loss functions whose gradients will drive the network to generate similar features for models within the same class while ensuring differences between models from different classes. 
Our primary approach involves leveraging two loss functions at different stages of the training procedure. 
The first one is a contrastive loss, which is very common for an unsupervised setup. 
The second one is a clustering loss, in which the clusters' centers are updated only every few epochs. This encourages the model to prioritize exploitation over exploration for short periods.

The contrastive loss moves each walk's features to get closer to the other augmentations of the same model but moves away from all other models. 
Specifically, we employ a contrastive loss known as {\em NT-Xnet,} introduced in SimCLR~\cite{chen2020simple}, and a novel clustering loss derived from the K-means algorithm. 
The contrastive loss aims to bring closer the features of two walks from the same model while moving them away from all other walks' features in the batch. 

The $NT\_Xnet$ loss is defined as follows:
\begin{equation}
    NT\_Xent Loss = -\frac{1}{2N} \sum_{i=1}^{N} \log \left( \frac{\exp(\text{sim}(x_i, x_i') / \tau)}{\sum_{j=1}^{2N} \exp(\text{sim}(x_i, x_j) / \tau)} \right)
    \label{eq:NTLoss}
\end{equation}
where \( N \) is the number of walks in a batch,
 \( x_i \) and \( x_i' \) are embeddings of a positive pair,
 \( \text{sim}(a, b) \) is the cosine similarity between vectors \( a \) and \( b \), and
 \( \tau \) is the temperature parameter.

While this loss may be effective in certain scenarios, it lacks the ability to retain knowledge between batches. The clustering loss, which is discussed hereafter, addresses and resolves this issue.

We propose to maintain clusters that are updated only from time to time so that they push the model to exploit a given configuration, rather than exploring a new one every batch. 
Each cluster should represent the set of features associated with a single class. 
We utilized the $K$-means algorithm for updating the cluster centers and the objective function. 
In our case, $K$ is the number of clusters, which is not necessarily the number of classes, as this is unknown in an self-supervised setting. The loss function of our $K$-means loss aims to minimize the within-cluster sum of squares (WCSS):
\begin{equation}
   KMeans\_Loss = \sum_{i=1}^{k} \sum_{j=1}^{n_i} \|x_j^{(i)} - \mu_i\|^2,
   \label{eq:Km-loss}
\end{equation}
where
\( x^{(i)} \) are all the feature vectors assigned to the mean \( \mu_i \).

There are three stages when using the clustering loss: Initialization, Assignment, and Means Update.

{\bf 1. Initialization:}
We initialize our clusters by placing the first cluster randomly and then choosing subsequent clusters with a higher probability of being far away from existing clusters.

{\bf 2. Assignment:}
Given the $i$-th positive pair of feature vectors, $(x^i_1, x^i_2)$, which are the features of two walks from the same model, we assign them both to a single mean, which is the closest to one of them. 
In particular, 
\begin{equation}
    \mu^i=argmin_{\mu\in U} (min({||x_1^i-\mu||^2}, {||x_2^i-\mu||^2}))
\end{equation}
where $U$ is the set of all means and $\mu^i$ is the chosen cluster for the $i$-th positive pair.
Ideally, there exists a single relevant cluster center for a particular class, making the closest mean a reasonable choice.

{\bf 3. Means update, as follows:}
From time to time, the means are updated by taking the mean of all feature vectors assigned to each cluster:
\begin{equation}
    \mu_i = \frac{1}{n_i} \sum_{j=1}^{n_i} x_j^{(i)},
\end{equation}
where $\mu_i$ is the mean to update and $n_i$ is the number of feature vectors assigned to this mean.

In practice, we initialize this loss with $80$ means and updated these means after every $5$ epochs. 
We use the clustering loss after $50$ epochs, as we primarily use this function to aid in exploitation, considering that most exploration has already been facilitated by the contrastive loss. 
We chose to use a $1:1$ ratio between the weight of the clustering loss and the contrastive loss when both are applied after $50$ epochs,
as follow:
\begin{equation}
    Loss = NT\_Xent Loss + \alpha *KMeans\_Loss
\end{equation}
In our case, $\alpha$ is set to $1$.

\subsection{Inference}
At the inference stage, our goal is to generate a feature vector for each given mesh. To achieve this, multiple walks are employed for the given mesh, with each walk producing a vector of features that represents the mesh. 
We assume that some walks may not capture unique aspects of the mesh, so we aim to eliminate these less informative walk features. Therefore, only half of the walks---those closest to the average of all walk feature vectors---are selected. 
These chosen walks are then averaged to produce the final feature vector representing the model.

To grasp the significance of walk selection and averaging, let's consider walks on a camel. Since walks are generated randomly, we anticipate that some will explore typical parts of the model, such as the body, which are similar to horse body. 
On the other hand, other walks are likely to explore unique parts, such as the hump or the head. The selective choosing and averaging process will most likely capture the features of a camel.

\section{Results}
The method of combining contrastive loss with clustering loss to extract qualitative features can be used for various downstream tasks, such as clustering and object retrieval.
We will evaluate the performance of our model on two downstream tasks: model retrieval and model clustering (classification). 

\subsection{Model Retrieval}
Given a query mesh, the goal is to accurately identify and retrieve similar meshes
based on the query mesh. 

For a given mesh, we generate multiple random walks, which are processed through the trained network. 
The network extracts a feature vector for each walk, and these vectors are then averaged into a single feature vector. 
Objects in the dataset are considered similar if their Euclidean distances to the given model are small.
In practice, we use $32$ walks per mesh model.

\paragraph{Datasets.}
To evaluate our model, we applied it to two datasets: SHREC11~\cite{veltkamp2011shrec} and ModelNet40~\cite{wu20153d}. These datasets vary in the number of classes, objects per class, and the types of shapes they include. 
SHREC11  consists of 30 classes, with 20 examples per class. 
Typical classes are camels, cats, glasses, centaurs, hands etc.
Modelnet40 contains 12, 311 CAD models from 40 categories, out of which 9,843 models are used for training and 2,468 models are used for testing. 
Unlike the SHREC11 dataset, many objects in this collection consist of multiple components and are not necessarily watertight, posing challenges for some mesh-based methods.


\paragraph{Evaluation metric.}
The mean Average Precision (mAP) is a common metric used to evaluate the performance of retrieval algorithms. It measures the quality of ranked retrieval results by calculating the average precision (AP) for each query and then averaging these values across all test queries. 
Given a set of queries, let \( Q \) be the total number of queries. For each query \( q \), the precision at a given rank \( k \), which is the number of retrieved elements, is defined as:
$$
\text{Precision}(k) = \frac{\text{Number of relevant items retrieved at rank } k}{k}
$$
The Average Precision (AP) for a query is the average of the precision values calculated at the ranks where relevant items are retrieved. It can be expressed as:
$$
\text{AP}(q) = \frac{1}{R} \sum_{k=1}^{N} \text{Precision}(k) \times \text{rel}(k)
$$
where \( R \) is the total number of relevant items for the query, \( N \) is the total number of retrieved items, and \( \text{rel}(k) \) is a binary function that equals 1 if the item at rank \( k \) is relevant, and 0 otherwise. The mean Average Precision (mAP) is then calculated as the mean of the AP scores across all queries:
$$
\text{mAP} = \frac{1}{Q} \sum_{q=1}^{Q} \text{AP}(q)
$$
This metric provides a single-figure measure of quality across recall levels, making it a popular choice for evaluating retrieval systems.

\paragraph{Results.}
Following the setup of ~\cite{lahav2020meshwalker, Hanocka_2019, Izhak2022AttWalk} , we split the objects in each class into 16 (/10) training examples and 4 (/10) testing examples. 

Figure \ref{fig:retrieval_shrec11} presents two examples of objects retrieved by our model for queries from the SHREC11 dataset: a bird and an ant. In both instances, the retrieved objects belong to the same class as the query, even when the birds and ants were in different poses. This indicates that our model has learned to recognize the general features of the classes, rather than simply memorizing specific instances.

\begin{figure}[t]
\centering
\includegraphics[width=0.45\textwidth]{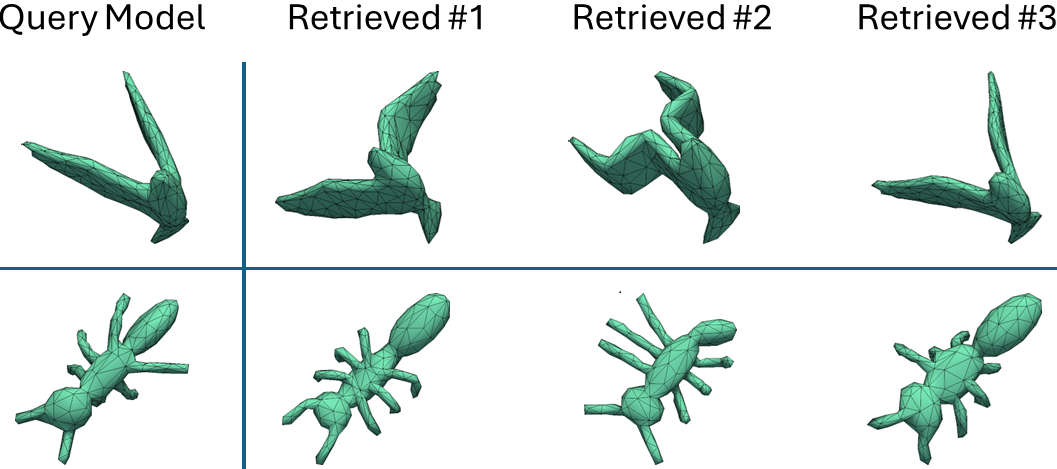}
\caption{ {\bf Object Retrieval.}
Given a query (on the left), our model retrieves objects from the same class as the query object, even when the pose (for the bird) or head orientation (for the ant) varies.
} 
\label{fig:retrieval_shrec11}
\end{figure}

Table \ref{shrec11_map} presents our performance on SHREC11. As our work is the first to be trained on this dataset in a self-supervised manner, we compare our results to those of supervised models for which results are available or could be trained, specifically MeshWalker~\cite{lahav2020meshwalker} and GWCNN~\cite{Ezuz2017GWCNN}.
For the two common split options (16/4 or 10/10), our self-supervised model achieves competitive results, performing nearly as well as the supervised models, with only a $2.7\%$ difference compared to the supervised approach.

\begin{table}[t]
\centering
\begin{tabular}{|l|c|c|}
\hline
Method & Split-16 & Split-10 \\
\hline
\hline
Ours (self-supervised) & 94.1\% & 89.8\% \\
\hline
\hline
MeshWalker~\cite{lahav2020meshwalker} (supervised) & 96.8\% & 94.3\% \\
\hline
GWCNN~\cite{Ezuz2017GWCNN} (supervised) & 96.0\% & 87.0\% \\
\hline
\end{tabular}
\caption{\textbf{Results (mAP) on SHREC11~\cite{veltkamp2011shrec}}. Split-16 and Split-10 are the number of training models per class (out of 20 models in the class). Our method achieves good results in both cases, comparable to those of the supervised MeshWalker and GWCNN models.
}
\label{shrec11_map}
\end{table}


Table \ref{modelnet40_map} presents the same experiment for ModelNet40. Once again, since there is no self-supervised work on ModelNet40, we compare our results to those of supervised models. In this instance, our results are below the performance of the supervised models. We hypothesize that this performance gap arises from the challenges self-supervised models encounter when
working with small datasets that exhibit high variability among objects within the same class.

\begin{table}[t]
\centering
\begin{tabular}{|l|c|}
\hline
Method & mAP score \\
\hline
\hline
Ours (self-supervised) & 71.3\%  \\
\hline
\hline
AttWalk~\cite{Izhak2022AttWalk} (supervised) & 91.2\% \\
\hline
MeshWalker~\cite{lahav2020meshwalker} (supervised) & 87.7\% \\
\hline
MeshNet~\cite{feng2018meshnetmeshneuralnetwork} (supervised) & 81.9\% \\
\hline
GWCNN~\cite{Ezuz2017GWCNN} (supervised) & 59.0\% \\
\hline
\end{tabular}
\caption{\textbf{Results on ModelNet40~\cite{wu20153d}}.
Supervised models outperform our self-supervised models, likely due to the high variability among objects within the same small class.
}
\label{modelnet40_map}
\end{table}


\subsection{SVM Classifier}

Another common task for evaluating self-supervised algorithms is training a linear SVM classifier over the learned features~\cite{crosspoint, achlioptas2018, wu2017}. The SVM classifier takes as input a set of feature vectors, typically represented as points in a high-dimensional space, and aims to find the optimal hyperplane that separates the classes with the maximum margin. The output of a linear SVM is a prediction, classifying each input instance into one of the classes. During training, the SVM learns the weights of the linear decision boundary, which can be interpreted as the importance of each feature in making the classification decision. Once trained, the SVM can efficiently classify new, unseen data points.

We process each mesh using our trained network, extracting feature vectors by averaging over multiple walks (as described in the model retrieval section). These feature vectors are then used as input for an SVM classifier. The dataset is split into training and test subsets, with the training set used to train the SVM and the test set for evaluation. To ensure robustness, we perform multiple runs and average the results.

To assess the performance of our SVM classifier, we applied it to the task of classifying meshes from two datasets: SHREC11~\cite{veltkamp2011shrec} and ModelNet40~\cite{wu20153d}. 
%
The classifier's performance is measured using accuracy, which represents the proportion of correct predictions across all instances. Accuracy is calculated by dividing the number of correctly classified samples by the total number of samples. In a multi-class classification task, accuracy reflects the overall correctness across all classes.

Figure \ref{fig:tsne_shrec11_10_classees} presents a t-SNE 2D visualization of meshes from 10 SHREC11 classes, along with their cluster centers. It demonstrates how effectively our features work, as a linear classifier can distinguish between the different classes.
Each class is grouped around a distinct cluster center, represented by a black star. Additionally, the 2D feature space generated by t-SNE reveals that semantically similar classes, such as 'dog1' and 'dog2,' are positioned close to each other, aligning with our expectations. 
Figure \ref{fig:tsne_modelnet40_10_classees} presents a t-SNE visualization of meshes from 10 classes of ModelNet40, along with their cluster centers. Similar to the previous figure, this one shows that classes which are visually similar to humans, such as tables and chairs, are positioned close to each other in the feature space.

However, while some classes are effectively grouped around distinct cluster centers, others show more noise and overlap with different centers. Notably, the 'desk,' 'dresser,' and 'night\_stand' classes experience significant confusion. Although this indicates imperfect performance by the linear SVM classifier, the confusion is understandable due to the considerable feature similarities among these classes.
For this dataset, our self-supervised model struggles more than a supervised approach to determine whether two samples from these similar classes belong to the same class. This difficulty may be attributed to class imbalances, which pose a challenge for contrastive models~\cite{Vito_2022}.

\begin{figure}[t]
\centering
\includegraphics[width=0.45\textwidth]{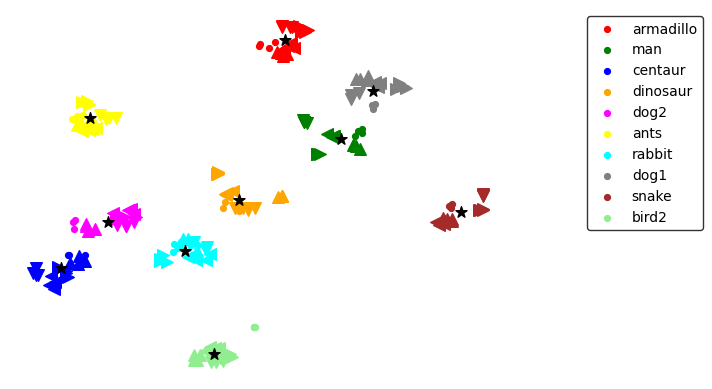}
\caption{ {\bf t-SNE of SHREC11 models features.}
This visualization shows the clustering of feature vectors extracted from 10 classes of the SHREC11 dataset. The plot demonstrates that models from the same class are grouped together. Notably, clusters representing similar classes, such as different dog breeds (e.g., 'dog1' and 'dog2'), are positioned close to each other in the feature space.
} 
\label{fig:tsne_shrec11_10_classees}
\end{figure}

\begin{figure}[t]
\centering
\includegraphics[width=0.45\textwidth]{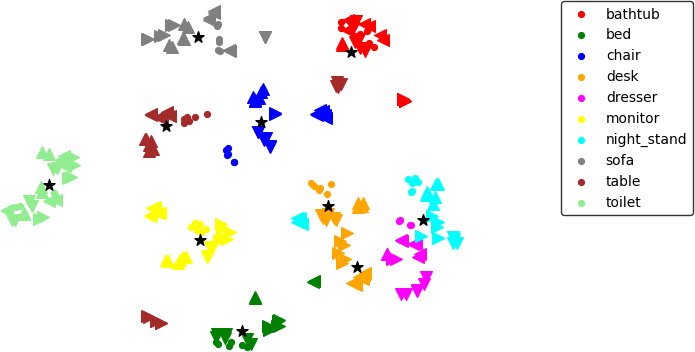}
\caption{ {\bf t-SNE of ModelNet40 models features.}
This figure shows the clustering of 10 ModelNet40 classes. Some classes, such as 'toilet', 'monitor', 'bathtub', are well-grouped. For others, where clustering is less ideal, the positioning is still logical, as similar classes are located close to each other in the feature space. For example, 'desk,' 'dresser,' and 'night\_stand' are positioned near one another.
} 
\label{fig:tsne_modelnet40_10_classees}
\end{figure}

Table \ref{svm_perf_shrec11} presents SVM classification accuracy for features extracted by our network and MeshWalker for the SHREC11 dataset. 
In both cases, these features were then used to train the linear SVM classifier, which we applied to the test data.
It shows a competitive results with a fully supervised model, with only a 2\% difference between our self-supervised model and the supervised MeshWalker classification accuracy using a linear SVM. 

Table \ref{svm_perf_modelnet40} presents our classification accuracy for ModelNet40, comparing it with the supervised MeshWalker method. Additionally, our evaluation framework allows us to compare our method against existing self-supervised approaches on ModelNet40 point clouds. We include these point cloud comparisons because, to our knowledge, there is no self-supervised work on ModelNet40 meshes. Our model's performance on ModelNet40 is 11.7\% lower than CrossPoint~\cite{crosspoint}. We believe this gap is due to our use of a contrastive approach in the loss function and challenges with unbalanced datasets.
 The issue arises because the number of 'False Negatives*' in each batch varies greatly depending on the class. Here, 'False Negatives*' refers to other objects from the same class that are treated as negatives because we don't use labels.

\begin{table}[t]
\centering
\begin{tabular}{|l|c|}
\hline
Method & SHREC11 Split-16\\
\hline
Ours (self-supervised) & 95.5\%  \\
\hline
\hline
MeshWalker~\cite{lahav2020meshwalker} (supervised) & 97.5\%  \\
\hline
\end{tabular}
\caption{\textbf{SVM classification on SHREC11~\cite{veltkamp2011shrec}.} 
The features extracted by our method yield competitive classification results when trained with a simple linear SVM classifier.
}
\label{svm_perf_shrec11}
\end{table}

\begin{table}[t]
\centering
\begin{tabular}{|l|c|c|}
\hline
Method &  Input & Modelnet40 \\
\hline
\hline
Ours (self-supervised)& Mesh & 79.5\%  \\
\hline
\hline
MeshWalker~\cite{lahav2020meshwalker} (supervised) & Mesh & 91.3\% \\
\hline
\hline
Crosspoint~\cite{crosspoint} (self-supervised) & Point Cloud & 91.2\% \\
\hline
Panos et al.~\cite{achlioptas2018} (self-supervised) & Point Cloud & 84.5\% \\
\hline
Jiajun et al.~\cite{wu2017} (self-supervised) & Point Cloud & 83.3\% \\
\hline
\end{tabular}
\caption{\textbf{SVM classification on ModelNet40~\cite{wu20153d}.} 
Our method is compared to SOTA results on ModelNet40 point clouds, as there are no comparable mesh works to our knowledge.}
\label{svm_perf_modelnet40}
\end{table}

\section{Conclusion}
This paper has introduced the first self-supervised learning system for 3D triangular meshes, addressing a gap in the field of 3D shape analysis. The key idea is to utilize random walks as a novel method of augmentation for 3D mesh models, enabling the extraction of high-quality features without the need for labelled data. This approach effectively tackles the challenges posed by the unordered and irregular nature of 3D meshes in deep learning algorithms.

Utilizing these random walks, the paper has proposed an end-to-end learning framework that leverages contrastive learning techniques and clustering loss. The random walks serve as a powerful mechanism to capture local geometric context, while the contrastive and clustering loss guides the learning process by maximizing similarity between augmented instances of the same mesh and minimizing similarity between instances from different meshes. This synergy allows our solution to uncover meaningful patterns and features in 3D mesh data.

Last but not least, the strength of this approach has been demonstrated through its applicability to downstream tasks. The extracted features have proven useful for two key applications in 3D shape analysis: object classification and shape retrieval.
Notably, our approach achieved excellent clustering results when evaluated on SHREC11, a small and balanced dataset, showing that the proposed method is promising. However, the results on ModelNet40 were less favorable, likely due to the nature of its classes and the imbalance between them.

While our results demonstrate the method's ability to learn discriminative features across a diverse range of 3D objects, underscoring its effectiveness in self-supervised learning tasks, future work should address the factors that limited its performance on ModelNet40. Additionally, the system should be explored in other 3D mesh-based tasks, especially in scenarios where labeled data is scarce or costly to obtain.

{\small
\bibliographystyle{ieee_fullname}
\bibliography{main}
}

\end{document}